\documentclass[useAMS,usenatbib,usegraphicx]{mn2e}

\newcommand{\der}{{\rm d}} 
 \newcommand{\mnras}{MNRAS}
\newcommand{\apj}{ApJ} 
\newcommand{\apjl}{ApJ} \newcommand{\nat}{Nature} 
\newcommand{\apjs}{ApJS}

\newcommand{\R}{{R_{\rm f}}} \newcommand{\wR}{\widetilde{R}_{\rm f}}

\newcommand{\co}{_{\rm c}} \newcommand{\cs}{_{\rm pk}}
 \newcommand{\tang}{_{\rm t}}
 
 \newcommand{\acc}{_{\rm acc}}
\newcommand{\pk}{_{\rm pk}} \newcommand{\nest}{^{\rm nest}}
 \newcommand{\T}{{\Re}} 
\newcommand{\fnest}{^{\rm 1st\,nest}}
\newcommand{\mini}{^{\rm min}} 
 \newcommand{\en}{_{\rm end}}
\newcommand{\cl}{_{\rm S}} \newcommand{\sub}{_{\rm sub}}
\newcommand{\tr}{^{\rm tr}} \newcommand{\dc}{_{\rm dc}} 
 \newcommand{\res}{_{\rm sim}} 

\newcommand{\p}{_{\rm p}} \newcommand{\rad}{_{\rm r}}
 \newcommand{\maxi}{_{\rm max}}
\newcommand{\modot}{M$_\odot$\ } \newcommand{\modotc}{M$_\odot$}
\newcommand{\ti}{t_{\rm i}}

 \newcommand{\nbody}{{$N$}-body }
\newcommand{\beq}{\begin{equation}} \newcommand{\eeq}{\end{equation}}
 \newcommand{\beqa}{\begin{eqnarray}}
\newcommand{\eeqa}{\end{eqnarray}} \newcommand{\lav}{\langle}
\newcommand{\rav}{\rangle} 
 \newcommand{\spot}{\lav\Phi\rav}
\newcommand{\srho}{\lav\rho\rav}

\begin{document}

\title[Substructure] {Theoretical dark matter halo substructure}

\author[Salvador-Sol\'e, Serra, \& Manrique]{Eduard
  Salvador-Sol\'e\thanks{E-mail: e.salvador@ub.edu}, Sinue Serra and
  Alberto Manrique\\ Institut
  de Ci\`encies del Cosmos, Universitat de Barcelona (UB--IEEC),
  Mart{\'\i} i Franqu\`es 1, E-08028 Barcelona, Spain}


\maketitle
\begin{abstract}
In two previous papers \citep{Sea12a,Sea12b}, it was
shown that: i) the typical structural and kinematic properties of
haloes in (bottom-up) hierarchical cosmologies endowed with random
Gaussian density perturbations of dissipationless collisionless dark
matter emerge naturally from the typical properties of peaks in the
primordial density field and ii) halo statistics are well described by
the peak formalism. In the present paper, we use these results to
model halo substructure. Specifically, making use of the peak
formalism and the fact that accreting haloes evolve from the
inside-out, we derive the subhalo mass abundance and number density
profile per infinitesimal mass for subhaloes of different masses, as a
function of the subhalo maximum circular velocity or mass, before and
after the tidal truncation of subhaloes by the host potential
well. The subhalo properties obtained by assuming that subhaloes are
mainly made of diffuse particles are in very good agreement with those
found in current high-resolution \nbody simulations. We also predict
the subhalo properties in the opposite extreme case, likely better
suited for the real universe in CDM cosmologies, that haloes are made
of subhaloes within subhaloes at all scales.
\end{abstract}

\begin{keywords}
methods: analytic --- galaxies: haloes --- cosmology: theory --- dark matter
--- haloes: substructure
\end{keywords}


\section{INTRODUCTION}\label{intro}

Halo substructure is attracting much attention for its implications in
the problem of galaxy formation. \citet{Kea99} and \citet{Moore99} 
noticed that numerical simulations of cold dark matter
(CDM) cosmologies found many more subhaloes at the galactic scale than
observed. Although the situation has notably changed since this
finding from the observational viewpoint 
(e.g. \citealt{WM97,Be06,Ko10,Bu10}), the
uncertainty remains on the consistency of CDM cosmologies with the
observed abundance of dwarf galaxies (e.g. \citealt{Sea10, Lea12}). 
Moreover, the distribution of subhalo maximum
circular velocities found in simulations has also been recently shown
to be in conflict with that observed in the Milky Way satellites
\citep{Boy11,VC11}.

The accurate determination of subhalo properties has been hampered for
a long time by the extremely high dynamic range required by \nbody
simulations addressing this issue. In the last decade, however, there
has been an impressive improvement (e.g. 
\citealt{Gea98,Spea01,Helmi02,Stea03,DMS04b,Kea04,DeL04,Gae04,Rea05}). 
The simulations by \citet{Dea07,Dea08} and \citeauthor{Spea08a} 
(\citeyear{Spea08b}, hereafter SWV) finally converged to a
well-determined subhalo mass function for Milky Way mass systems in
the concordance cosmology. More recent simulations with similar
resolutions have begun to study subhalo properties in host haloes
with different masses, redshifts and concentrations \citep{ALBF09,EWT09,Kea11,Gae11}.

The situation is less satisfactory, however, from the theoretical
viewpoint. The origin of substructure is well-understood: it is the
consequence of the halo hierarchical growth and the fact that, when haloes
are captured by a more massive object, they are not fully destroyed by
the host tidal field. But the details of the process are not
clear enough and the models developed so far (e.g. \citealt{Fea02,S03,OL04}) 
are still unable to recover the results of
\nbody simulations. This is embarrassing not only because of the need
to fully understand this phenomenon, but also because, due to their
high cost in CPU time, current \nbody simulations with still rather
limited dynamic range inform us on substructure in haloes with only a few
masses and redshifts.

In two recent papers, \citeauthor{Sea12a}~(\citeyear{Sea12a,Sea12b}, hereafter
SVMS and SSMG, respectively) have shown that the smooth structural and
kinematic properties of haloes (related to the 1-particle probability
distribution function) endowed with dissipationless collisionless dark
matter emerge from the properties of peaks in the density field at an
arbitrarily small cosmic time, determined by the power-spectrum of
density perturbations in the (bottom-up) hierarchical cosmology under
consideration. This allowed these authors to justify the peak
formalism (\citealt{D70,DS78,BBKS}, hereafter BBKS; \citealt{PH90,BM96}), 
more specifically, the rigorous version of it developed by
\citeauthor{MSS95} (\citeyear{MSS95,MSS96}, hereafter MSSa and MSSb; see
also \citealt{metal98}) and show that this is a very useful tool
to deal with halo statistics.

In the present paper, we use this formalism and the model for the
smooth structural and kinematic properties of haloes developed in SVMS
and SSMG to describe the properties of CDM halo substructure (related
to the $n$-particle probability distribution function, with $n\ge
2$). Although \nbody simulations are nowadays able to study
clumps-in-clumps (e.g. SWV), the only complete studies carried out so
far on the properties of halo substructure concern {\it first-level}
subhaloes. For this reason, we focus on this particular kind of
substructure, although we also draw conclusions on their internal
structure. Specifically, we derive the subhalo abundance and number
density profile per infinitesimal mass, as a function of subhalo
maximum circular velocity and mass, before and after the tidal
truncation of subhaloes by the host potential well. Although the
formalism is developed under the assumption that haloes form by pure
accretion (PA), the results obtained are shown to be also valid for
haloes having undergone major mergers. The theoretical predictions
obtained for the CDM concordance model with $(\Omega_{\rm
  m},\Omega_\Lambda,h,\sigma_8)=(0.3,0.7,0.7,0.9)$ are in very good
agreement with the results of numerical simulations.

The paper is organised as follows. In Section \ref{peaks}, we recap
the main results obtained in SVMS, SSMG, MSSa and MSSb in connection
with the present study. These results are used in Section
\ref{abundance} and Section \ref{numdens} to derive the subhalo
abundance as a function of maximum circular velocity and the number
density of subhaloes per infinitesimal original (non-truncated) mass,
respectively. The correction for tidal truncation is given in Section
\ref{truncation}. In Section \ref{friction}, we discuss the effects of
dynamical friction. Our results are summarised in Section \ref{dis}. A
package with the numerical codes used in the present paper is publicly
available from {\texttt {www.am.ub.es/$\sim$cosmo/haloes\&peaks.tgz}}.

\section{THEORETICAL BASIS}\label{peaks}

As shown in SVMS, in accretion periods between consecutive major
mergers haloes develop outwardly by keeping their instantaneous inner
region unaltered. As a consequence of this growth, the radius $r$
encompassing the mass $M$ in a (triaxial) virialised halo grown by PA
(i.e. having suffered no major merger) from a (triaxial) protoobject
with outward-decreasing spherically averaged density profile and
Hubble flow-dominated kinematics at an arbitrarily small time
$\ti$ (where the protohalo is in linear regime) satisfies the
relation
\beq 
\frac{3}{10}\frac{r|{\cal E}\p(M)|}{GM^2}=1\,,
\label{vir2}
\eeq
where ${\cal E}\p(M)$ is the total energy within the sphere with mass
$M$ of the spherically averaged protohalo. Equation (\ref{vir2})
allows one to infer the mass profile $M(r)$ of the virialised halo
and, hence, its spherically averaged density profile $\srho(r)$. 

In SSMG, it was shown that the relation
\beq 
\beta(r)=
1-\frac{\sigma\tang^2(r)}{\sigma\rad^2(r)}=1-\frac{\frac{\sigma\tang^2(r)}{\sigma^2(r)}}{1-2\frac{\sigma\tang^2(r)}{\sigma^2(r)}}\,,
\label{beta}
\eeq
with
\beq 
\frac{\sigma^2\tang(r)}{\sigma^2(r)}=\left\lav\!\left(\frac{\delta\Phi}{\spot}\right)^{\!2}\right\rav^{\!1/2}\!\!\!(r)\,.
\label{00th}
\eeq
between the anisotropy profile, $\beta(r)$, and the rms scaled
potential fluctuation profile, $\lav (\delta\Phi/\spot)^2\rav^{1/2}(r)$,
related to the same profile in the protohalo, $\lav
(\delta\Phi\p/\ \lav\Phi\p\rav)^2\rav^{1/2}(r\p)$, and the generalised
Jeans equation for anisotropic triaxial systems (see SVMS),
\beq
\frac{\der [\srho\,\sigma^2/(3-2\beta)]}{\der r}\!+\!
\frac{2\beta(r)}{r}\frac{\srho(r)\sigma^2(r)}{3-2\beta(r)}
\!=\!-\srho(r)\frac{GM(r)}{r^2}\!,
\label{exJeq2}
\eeq
for the velocity variance, $\sigma^2(r)$, allow one to derive the
$\beta(r)$ and $\sigma(r)$ profiles for haloes from the triaxial shape
of peaks acting as their putative seeds by PA.

The typical properties of peaks, namely the quantities ${\cal E}\p(M)$
and $\lav (\delta\Phi\p/\ \lav\Phi\p\rav)^2\rav^{1/2}(r\p)$, in the
primordial density field were derived from the power-spectrum of
density perturbations making use of the peak formalism. This formalism
is based on the peak Ansatz stating that there is a one-to-one
correspondence between haloes and peaks as suggested by
PA. Specifically, any halo with $M$ at the time $t$ is traced by a
peak in the density field at $\ti$ filtered by means of a Gaussian
window, with density contrast $\delta\pk$ and filtering scale $\R$,
given by
\beq 
\delta\cs(t)=\delta\co(t) \frac{G(\ti)}{G(t)}\,~~~~~~~~
\R(M)=\frac{1}{q}\left[\frac{3M}{4\pi\bar\rho(\ti)}\right]^{1/3}\!\!,
\label{rm}
\eeq
where $G(t)$ is the cosmic growth factor, $q$ is the radius, in units
of $\R$, of the collapsing cloud with volume equal to $M$ over the
mean cosmic density at $\ti$, $\bar\rho(\ti)$, and $\delta\co(t)$ is the
critical linearly extrapolated density contrast for current
collapse. For the concordance model, the best values of $q$ and
$\delta\co(z)$ are respectively 2.75 and $1.93+(5.92-0.472 z+0.0546
z^2)/(1+0.000568 z^3)$, where $z$ is the redshift corresponding to the
cosmic time $t$.

As haloes accrete, their associated peaks describe continuous
trajectories in the $\delta\cs$--$\R$ diagram. (The peaks over the
trajectory are {\it connected} between each other in the sense
explained in SSMa; see also SVMS.) The typical peak trajectory leading
to a halo with $M_0$ at $t_0$ is the solution of the differential
equation,
\beqa
\frac{\der \delta\cs}{\der \R}=\left\{r_{{\rm mass}}^{{\rm a}}[M(\R),t(\delta\cs)]\,\frac{\der \R}{\der M}\,\frac{\der t}{\der \delta\cs}\right\}^{-1}\nonumber\\
= -x_{{\rm e}}(\delta\cs,\R)\,\sigma_2(\R) \R\,,~~~~~~~~~~~~~~~~~
\label{dmd}
\eeqa
for the boundary condition $\delta\pk(t_0)$ at $\R(M_0)$, where
$x_{{\rm e}}(\delta\cs,\R)$ is the inverse of the average (close to
the most probable) inverse curvature $x$ (equal to minus the Laplacian
over $\sigma_2$),
\beq \left\lav
\frac{1}{x}\right\rav (\R,\delta\cs)\!=\! {(2\pi)^{-1/2} \over
  (1-\gamma^2)^{1/2}}\!\! \int_0^\infty\!\! \der x\,\frac{1}{x}\,
f(x)\,{\rm e}^{-{(x-x_\star)^2 \over 2(1-\gamma^2)}}\,,
\label{G}
\eeq
in peaks with $\delta\cs$ and $\R$ (BBKS), being
\beqa
f(x)=\frac{x^3-3x}{2}\left\{{\rm erf}\!\left[\left(\frac{5}{2}\right)^{1/2}x\right]+{\rm erf}\!\left[\left(\frac{5}{2}\right)^{1/2}\frac{x}{2}\right]\right\}\nonumber~~\\
+\left(\frac{2}{5\pi}\right)^{\!\!1/2}\!\!\left[\left(\!\frac{31x^2}{4}+\frac{8}{5}\!\right){\rm e}^{-\frac{5x^2}{8}}+\left(\!\frac{x^2}{2}-\frac{8}{5}\!\right){\rm e}^{-\frac{5x^2}{2}}\right]\!,
\label{fx}
\eeqa
where $\gamma$ and $x_\star$ are respectively defined as
$\sigma_1^2/(\sigma_0\sigma_2)$ and $\gamma \delta\cs/\sigma_0$ in
terms of the j-th order spectral moments for the power-spectrum
$P(k)$,
\beq
\sigma_j^2(\R)=\int_{0}^\infty \frac{\der k}{2\pi^2} \,P(k)\,k^{2j+2}\,{\rm e}^{-k^2\R^2}\,.
\label{spec}
\eeq
The first equality in equation (\ref{dmd}) relates the derivative of
$\delta\pk(t)$ to the typical mass accretion rate, $r_{{\rm
    mass}}^{{\rm a}}(M,t)$, of haloes with $M$ at $t$ (see MSSb).

When a halo suffers a major merger, the $\delta\pk(\R)$ trajectory of
its associated peak is {\it interrupted} (there is no peak at the
immediately larger scale to be connected with). At the same time, a
new peak appears (there is no peak at the contiguous smaller scale to
be connected with), with the same density contrast as the disappeared
peak but at a substantially larger scale, that traces the new halo
formed in the major merger. Major mergers are the only way peak
trajectories are interrupted. When a halo is accreted by another much
more massive halo, the trajectory of the associated peak is not
interrupted, so the peak becomes nested into the collapsing cloud of
the larger scale peak with identical density contrast tracing the
accreting halo. On the other hand, when a peak disappears because the
associated halo suffers a major merger, its trajectory is interrupted
but not those of its nested peaks, which also become nested in the
collapsing cloud of the new peak resulting from the merger. In this
way, a complex system of peak nesting at multiple levels is built
similar to the nesting of subhaloes in haloes (see
Sec.~\ref{abundance}).

One important implication of the SVMS and SSMG model is that the
properties of haloes having suffered major mergers are
indistinguishable from those of haloes grown by PA. This result allows
one to understand why the typical density and kinematic profiles for
haloes grown by PA are representative of all haloes, regardless of their
aggregation history, and why the peak Ansatz suggested by PA also
holds for all haloes, regardless of their actual aggregation history.

Below, we detail the form of several unconditioned and conditional
number densities of peaks, nested or non-nested within other peaks
calculated in MSSa, MSSb and \citet{metal98} that will be used
in the following sections. Note that all these quantities depend on
the power-spectrum of the hierarchical cosmology considered through
the spectral moments defined above.

The number density of peaks with density contrast $\delta\pk$ at
scales $\R$ to $\R + \der\R$ is the number density of peaks at
scale $\R$ with density contrast $\delta$ greater than $\delta\pk$
that cross such a density contrast when the scale is increased to $\R
+ \der\R$ or, equivalently, with $\delta$ satisfying the condition
\beq
\delta\pk < \delta \le \delta\pk+\sigma_2(\R)\,x\,\R\,\der\R\,.
\label{cond}
\eeq 
Thus, such a number density can be obtained by integrating
over $\delta$ and $x$ the density of peaks with height
$\nu=\delta/\sigma_0(\R)$ and curvature $x$ in infinitesimal ranges,
${\cal N}\pk(\nu,x,\R)\,\der\nu\,\der x$, calculated by BBKS,
\beqa
N\pk(\R,\delta\pk)\,\der \R=\!\!\int_0^\infty\!\!\!\!\der x\int_{\nu\pk}^{\nu\pk'}\!\!\der \nu \,{\cal N}\pk(\nu,x,\R)\nonumber\\
=\frac{\lav x\rav(\R,\delta\pk)}{(2\pi)^2R_\star^3}\,\,
{\rm e}^{-\frac{\nu\pk^2}{2}}\,
{\sigma_2(\R)\over\sigma_0(\R)}
\,\R\,\der\R~~~~~~
\label{npeak}
\eeqa
where $\nu\pk'=\nu\pk+[\sigma_2(\R)/\sigma_0(\R)]\R\der\R$ and $\lav
x\rav$ is the average curvature of peaks with $\delta\pk$ and $\R$
\beq
\lav x\rav(\R,\delta\pk)\!=\! {(2\pi)^{-1/2} \over (1-\gamma^2)^{1/2}}\!\! \int_0^\infty\!\! \der x\,x\,
f(x)\,{\rm e}^{-{(x-x_\star)^2 \over 2(1-\gamma^2)}}\,.
\label{oldG}
\eeq

Likewise, the conditional number density of peaks with $\delta$ at
scales $\R$ to $\R + \der\R$ subject to being located in the
collapsing cloud of a non-nested background peak with $\delta'$ at
$\R'$, $N\pk\nest(\R,\delta|\R',\delta')$ $\der\R$, can be obtained by
integrating the conditional number density of peaks with $\delta$ at
scales $\R$ to $\R + \der\R$ subject to being located at a distance
$r$, in units of $q\R$, from the background peak,
$N\pk(\R,\delta|\R',\delta',r)$, out to $q=1$,
\beqa N\pk\nest(\R,\delta|\R',\delta')=
C^{-1}\!\!\!\int_0^1\!\!
\der r\, 3 r^2 N\pk(\R,\delta|\R',\delta',r)\,,
\label{int}
\eeqa
with the latter conditional number density of peaks obtained, as the
ordinary number density above, by integrating over $\nu$ and $x$ the
conditional density of peaks with those arguments in infinitesimal
ranges, subject to being located at the distance $r$ from a
background peak with $\nu'$ at $\R'$, ${\cal
  N}\pk(\nu,x,\R|\R',\delta',r)\,\der \nu\,\der x$, calculated by
BBKS, with $\delta$ satisfying the condition (\ref{cond}),
\beqa 
N\pk(\R,\delta\pk|\R',\delta',r)\der \R\!=\!\!\!\!\int_0^\infty\!\!\!\!\!\!\der
x\!\!\!\int_{\nu\pk}^{\nu\pk'}\!\!\!\der \nu{\cal N}\pk(\nu,x,\R|\R'\!,\delta'\!,r),
\nonumber
\eeqa
\vskip -10pt
\beqa
\label{int2}
~~~~~=\frac{\lav x\rav(\R,\delta\pk,r)}{(2\pi)^2\,R_{\star}^3\,e(r)}
      {\rm e}^{-\frac{\left[\nu\pk -
            \epsilon(r)\,\nu'(r)\right]^2}{2e(r)^2}}{\frac{\sigma_2(\R)}{\sigma_0(\R)}}
      \,\R\,\der\R 
\eeqa 
where $\lav x\rav(\R,\delta\pk,r)$ is the average curvature of peaks
with $\delta\pk$ and $\R$ at a distance $r$ from a peak, given by
\beq
\lav x\rav(\R,\delta\pk,r)\!=\!
\frac{(2\pi)^{-1/2}}{[1-\tilde\gamma^2(r)]^{1/2}}\!\! \int_0^\infty \!\!\!\der x\,
x\,f(x)\,{\rm e}^{-{[x-\tilde x_\star(r)]^2 \over 2[1-\tilde\gamma^2(r)]}}\!,.
\label{mf17}
\eeq
In equation (\ref{mf17}), we have used the following notation: $e(r)=\sqrt{1
- \epsilon(r)^2}$, $\tilde x_\star(r) =
\tilde\gamma(r)\,\tilde\nu(r)$, $\tilde\gamma^2(r) =
\gamma^2\left[1 + \epsilon(r)^2\,{(1 - r_1)^2 \over 1 -
\epsilon(r)^2}\right]$, $r_1 = \left({\R/ R_h}\right)^2$,
$R_h^2=(\R^2+\R'^2)/2$,
$\epsilon(r)\!=\!\!\left(\R \R'/R_h^2\right)^{(n+3)/2}g(r,\R')$,
$\nu'(r)=\frac{\overline{\delta'(r)}}{\sigma_0(\R')}\,\,g(r,\R')$,
$g(r,\R')=\left\{1-[\Delta\delta'(r)]^2/\sigma_0(\R')\right\}^{1/2}$
and
\vskip -10pt
\beqa
\tilde\nu(r)\!=\!{\gamma \over \tilde\gamma(r)}\,{1 - r_1 \over 1 -
\epsilon(r)^2}\left[\nu\!\left({1 - \epsilon(r)^2r_1 \over 1 - r_1}\!
\right)\!-\epsilon(r)\nu'(r)\right]\!,\nonumber
\label{tild}
\eeqa
\vskip -3pt
\noindent with the mean and rms density contrast at $r$ from a peak,
$\overline{\delta'(r)}$ and $[\Delta\delta'(r)]^2$, respectively equal
to
\beqa
\overline{\delta'(r)}=\frac{\gamma\delta}{1-\gamma^2}\left(\frac{\psi}{\gamma}+\frac{\nabla^2\psi}{u^2}\right)-\frac{x\sigma_0(\R)}{1-\gamma^2}\left(\gamma\psi+
\frac{\nabla^2\psi}{u^2}\right)
\label{e1}
\eeqa
\beqa
[\Delta\delta'(r)]^2=\sigma_0^2\bigg\{\!1\!-\!\frac{1}{1-\gamma^2}\left[\psi^2+\!\!\left(2\gamma\psi+\frac{\nabla^2\psi}{u^2}\!\!\right)\frac{\nabla^2\psi}{u^2}\right]\nonumber\\
-5\left(\frac{3\psi'}{u^2r}-\frac{\nabla^2\psi}{u^2}\right)^2-\frac{3(\psi')^2}{\gamma u^2}\bigg\}\,,~~~~~~~~~~~~~~~~~~~~
\label{e2}
\eeqa
being $\psi$ the ratio $\xi(r)/\xi(0)$, $\psi'$ its $r$-derivative,
$\xi(r)$ the mass correlation function at $\R$ and $u$ equal to
$(q\R)^2\sigma_2(\R)/\sigma_0(\R)$. In equation (\ref{int}), the
factor
\beqa
C\equiv\frac{4\pi s^3 N(\R',\delta')}{3N\pk(\R,\delta)}
\int_0^{s}\der r\,3 r^2\,N\pk(\R,\delta|\R',\delta',r)\,,
\eeqa
with $s$ equal to the mean separation between non-nested peaks drawn
from their mean density (eq.~[\ref{nnp}] below), is to correct for the
overcounting of background peaks, as in the preceding calculation of
$N\pk(\R,\delta|\R',\delta',r)$ they are not necessarily non-nested.

Finally, given the preceding number densities, it is readily seen that
the number density of {\it non-nested} peaks with $\delta$ at scales
$\R$ to $\R + \der\R$, $N(\R,\delta)\der \R$, is the solution of the
Volterra integral equation correcting the ordinary number density of
peaks (eq.[\ref{npeak}]) for nesting,
\beqa
N(\R,\delta)=N\pk(\R,\delta)\nonumber~~~~~~~~~~~~~~~~~~~~~~~~~~~~~~~~~~~~~~~~~~~\\
-\frac{1}{\bar\rho(\ti)}\int_\R^\infty d\R' M(\R')\,N(\R,\delta)\,N\pk\nest(\R,\delta|\R',\delta),
\label{nnp}
\eeqa
and that the number density of peaks with $\delta\pk$ at scales
$\R$ to $\R+\der \R$, {\it nested in non-nested peaks} with $\delta'\pk$ at
scales $\R'$ to $\R'+\der \R'$, is given by
\beqa 
N\nest(\R\rightarrow \R',\delta\pk\rightarrow \delta'\pk)\der \R\,\der
\R'\nonumber~~~~~~~~~~~~~~~~~~~~~~~~\\
=N\pk\nest(\R,\delta\pk|\R',\delta'\pk)\,\der \R N(\R',\delta'\pk)\,\frac{M(\R')}{\bar\rho(\ti)}\,\der \R'\!,
\label{arrow2}
\eeqa
where $M(\R')/\bar\rho(\ti)$ is the volume of the collapsing cloud of
the peak with $\delta'\pk$ at $\R'$ and
$N\pk\nest(\R,\delta\pk|\R',\delta'\pk)$ is the conditional number
density of peaks with $\delta\pk$ at $\R$ subject to being located in
the collapsing cloud of non-nested peaks with $\delta'\pk$ at $\R'$.

\section{SUBHALO ABUNDANCE}\label{abundance}

In dark matter clustering, first-level subhaloes develop in two ways:
i) through the accretion by a halo of much less massive partners (with
substantially higher concentrations), which become first-level
subhaloes of the accreting halo at the same time that their own
first-level clumps become second-level ones and so on; and ii) through
major mergers of similarly massive haloes (with similar
concentrations), where the merging objects meld and their respective
first-level subhaloes are transferred as such to the new halo
resulting from the merger.

As explained in Section \ref{peaks}, the processes of halo accretion
and major mergers are correctly traced by peak trajectories in the
$\delta\pk$--$\R$ diagram. Furthermore, the halo-nesting they produce
is also correctly traced by the corresponding peak-nesting. Indeed,
when haloes are accreted, the peaks tracing them survive and become
nested into the collapsing clouds of those peaks tracing the accreting
haloes, while the peaks already nested within them become second-level
nested peaks and so on. On the other hand, in major mergers, peaks
tracing the merging haloes disappear and their first-level
(second-level,...)  nested peaks automatically become so in the
collapsing cloud of the new peak tracing the halo formed in the
merger. Both behaviours reproduce that above mentioned of haloes and
subhaloes in accretion and major mergers. Thus, by counting the
first-level peaks nested in the collapsing cloud of peaks in the
density field at $\ti$, we can estimate the number of first-level
subhaloes in the associated haloes at $t$.

The total number of first-level peaks with $\delta\pk$ and scales
greater than $R\cl$ nested within the collapsing cloud of a non-nested
peak with $\delta\pk$ at scale $\R$, $N(>
R\cl,\delta\pk|\R,\delta\pk)$, follows from equation
(\ref{arrow2}), the result being
\beqa 
\!N(>\!\!
R\cl,\delta\pk|\R,\delta\pk\!)\!\!=\!\!\frac{M}{\bar\rho(\ti)}\!\!\int_{R\cl}^{\R}
\!\!\!\der \wR \,\!\bigg\{\!
N\pk\nest(\wR,\delta\pk|\R,\delta\pk)\nonumber
\eeqa
\vskip -10pt
\beqa
~~~~~~~-\int_{{\wR}}^{\R}\!\!
\der
\R'\,N\pk\fnest(\wR,\delta\pk|\R',\delta\pk)\nonumber\\ \times
N\pk\nest(\R',\delta\pk|\R,\delta\pk)\,\frac{M(\R')}
{\bar\rho(\ti)}\!\bigg\}.~\label{peaks2} \eeqa
In equation (\ref{peaks2}), the integral over $\R'$ is to correct the
number density of peaks nested in the seed of the halo for those peaks
nested in intermediate-scale peaks so as to ensure that only
first-level nested peaks are counted. The factor
$N\pk\nest(\R',\delta\pk|\R,\delta\pk)$ inside that integral comes
from the probability for the intermediate peaks to be nested in the
seed of the halo, equal to
$N\pk\nest(\R',\delta\pk|\R,\delta\pk)N(\R,\delta\pk)/N\pk(\R',\delta\pk)$
times $M/\bar\rho(\ti)$, and the function
$N\pk\fnest(\wR,\delta\pk|\R',\delta\pk)$, solution of the Volterra
integral equation
\beqa 
N\pk\fnest(\wR,\delta\pk|\R',\delta\pk)\equiv N\pk\nest(\wR,\delta\pk|\R',\delta\pk)
\nonumber\\
-\int_{\wR}^{\R'} \der \R''\,N\pk\nest(\wR,\delta\pk|\R'',\delta\pk)\nonumber~~~~~~~~~\\
\times N\pk\fnest(\R'',\delta\pk|\R',\delta\pk)\frac{M(\R'')}
{\bar\rho(\ti)}\,,~~~~~~
\label{cor}
\eeqa
gives the conditional number density of peaks with $\delta\pk$ at scale
$\wR$ subject to reach {\it for the first time} the same density
contrast at an intermediate scale $\R'$ when the scale is increased
from $\wR$. This ensures that the correction for intermediate
nesting is not overcounted. 

\begin{figure}
\vskip -13pt
\centerline{\includegraphics[scale=0.46]{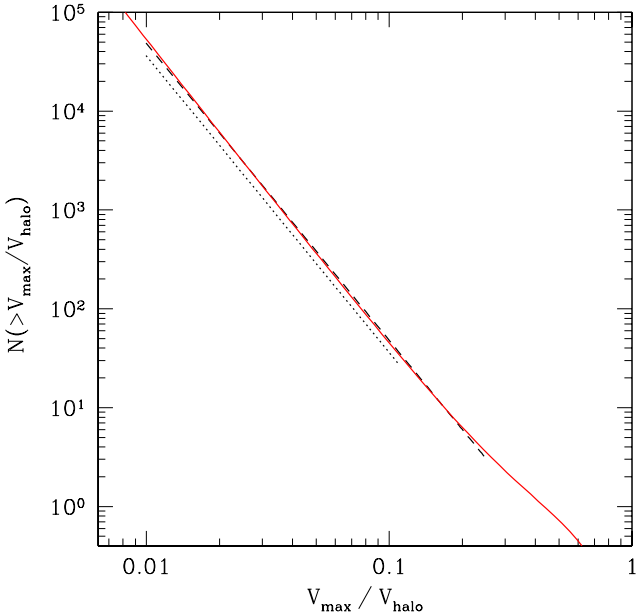}}
\caption{Theoretical cumulative abundance of non-truncated subhaloes (solid
red line) for a Milky way mass halo in the concordance cosmology as a
function of subhalo maximum circular velocity, $V\maxi$, scaled to
that of the halo. For comparison, empirical curves obtained by
SWV (dashed black line) and \citet{Dea08} (dotted black line).}
\label{f1}
\end{figure}

\begin{figure}
\vskip -13pt
\centerline{\includegraphics[scale=0.46]{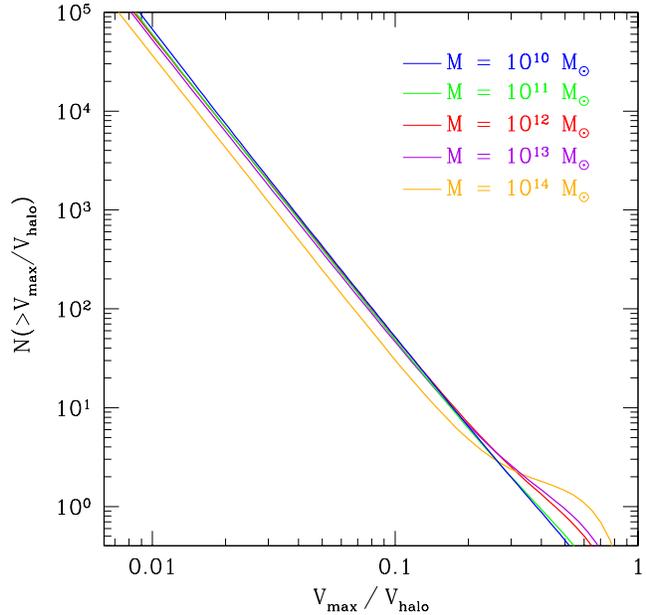}}
\caption{Same as Figure \ref{f1} but for several halo masses (coloured
lines).}
\label{f2}
\vskip 10pt
\end{figure}

As first-level subhaloes with masses greater than $M\cl$ in a halo
with $M$ at $t$, $N(>M\cl,t)$ are correctly traced by first-level
peaks with $\delta\pk(t)$ at scales greater than $R\cl=\R(M\cl)$
(eqs.~[\ref{rm}]) nested in the collapsing cloud of a non-nested peak
with $\delta\pk(t)$ and $\R(M)$, their cumulative abundance
$N(>M\cl,t)$ must also be well-estimated by the corresponding
abundance of nested peaks, $N(> R\cl,\delta\pk|\R,\delta\pk)$, given
by equation (\ref{peaks2}). In Figure \ref{f1}, this theoretical
cumulative subhalo abundance for current Milky Way mass ($1.4\times
10^{12}$ \modotc) haloes in the concordance cosmology is compared to
that found in numerical simulations by \citet{Dea08} and
SWV\footnote{The halo mass is $M_0=1.4\times 10^{12}$ \modot in all
  cases, although the mass is defined within $r_{50}$ in \citet{Dea08}, 
  $r_{200}$ in SWV and $\sim r_{90}$ in the present paper
  (see above). Nonetheless, all three curves overlap at the scale of
  Figure \ref{f1}.}. As can be seen, there is excellent agreement,
particularly in the case of SWV results. Figure \ref{f2} shows the
theoretical subhalo mass abundance as a function of the scaled maximum
circular velocity for different halo masses. Except for a small shift
at the large mass end, the predicted subhalo abundance is essentially
independent of halo mass, in agreement with a very common idea, though
with rather limited empirical support.

As can be seen, the theoretical cumulative subhalo abundance shows a
small bump at large masses which arises from a similar bump in the
conditional peak number density,
$N\pk\nest(\wR,\delta\pk|\R,\delta\pk)$, at scales $\wR$ comparable to
$\R$. This latter function is approximate for $\wR$ close to $\R$ (see
\citealt{metal98}): it should vanish when $\wR$ approaches $\R$
more rapidly than it actually does\footnote{Not only can there be no
  peaks nested in other peaks with identical scale but also within
  peaks with slightly larger scale. The capture by a halo of another
  similarly massive one necessarily causes a major merger, so the two
  peaks disappear.}. This suggests that these bumps may be an artefact
due to the less steep fall of $N\pk\nest(\wR,\delta\pk|\R,\delta\pk)$
at large scales. But this is hard to ascertain. Empirical data are too
noisy there to asses the reality or not of the bump in the subhalo
abundance. In fact, there are indications that it is real: had we only
slightly sanded the bump in the conditional peak number density, the
resulting subhalo abundance would take negative values. For this
reason, we have preferred to conserve it and adopt a sharp cutoff at
$\wR/\R$ equal to one tenth for subhalo masses greater than one
hundredth of the host mass and at $\wR/\R$ equal to one hundredth
otherwise. Such a cutoff does not essentially alter the theoretical
subhalo abundance shown in Figures \ref{f1} and \ref{f2} while it
notably improves the behaviour of the subhalo number density profile
derived below for subhalo masses close to the host mass.


\section{SUBHALO NUMBER DENSITY PROFILE}\label{numdens}

Given the inside-out growth of haloes formed by PA (see
Sec.~\ref{peaks}), the cumulative abundance of subhaloes with masses
greater than $M\cl$ within the radius $r$, $N(<\! r,\, >\! M\cl)$,
must coincide with the cumulative subhalo abundance by the time the
halo radius was equal to $r$. Consequently, the differential subhalo
abundance, both per infinitesimal halo radius and subhalo mass, in a
halo with $M_0$ at $t_0$ is
\beq 
N(r,M\cl)\!=\!\!\left(\!\!\!\left\{\!\!\left[\frac{\der N_{\delta\pk}(M\cl)}{\der \delta\pk}\!\right]_{\!\delta(\R)}\!\!\!\frac{\der\delta\pk}{\der \R}\!\right\}_{\!\!\R(M)}\!\!\!\frac{\der \R}{\der M}\!\right)_{\!\!\!M(r)}\!\!\!\!\!\frac{\der M}{\der r},
\label{densclump} 
\eeq
where 
\beqa 
N_{\delta\pk}\!(M\cl)\!=\!\!\frac{M[\R(\delta\pk)]}{D(M\cl)}
\bigg\{\!N\pk\nest\left[\R(M\cl),\delta\pk|\R(\delta\pk),\delta\pk\right]
\nonumber
\eeqa
\vskip -7pt
\beqa
~~~~~-\int_{\R(M\cl)}^{\R(\delta\pk)}
\der \R'\,N\pk\fnest[\R(M\cl),\delta\pk|\R',\delta\pk]\nonumber~~~~~~\\
\times N\pk\nest[\R',\delta\pk\left|\R(\delta\pk),\delta\pk\right]\frac{M(\R')}{\bar\rho(\ti)}\bigg\},~~~~~~~~~~~~~
\label{densclump2} 
\eeqa
is the differential subhalo abundance obtained by differentiation of
the cumulative subhalo abundance $N(> M\cl)$, given in equation
(\ref{peaks2}) for $M$ and $t$ respectively equal to
$M[\R(\delta\pk)]$ and $t(\delta\pk)$, with $D(M\cl)=q\left\{\pi\left[6\bar\rho^2(\ti) M\cl\right]^2\right\}^{1/3}$, $M(r)$ given by equation
(\ref{vir2}), $\R(M)$ given by equation (\ref{rm}) and $\R(\delta\pk)$
equal to the inverse typical peak trajectory solution of equation
(\ref{dmd}) leading to a halo with $M_0$ at $t_0$.

\begin{figure}
\vskip -15pt
\centerline{\includegraphics[scale=0.46]{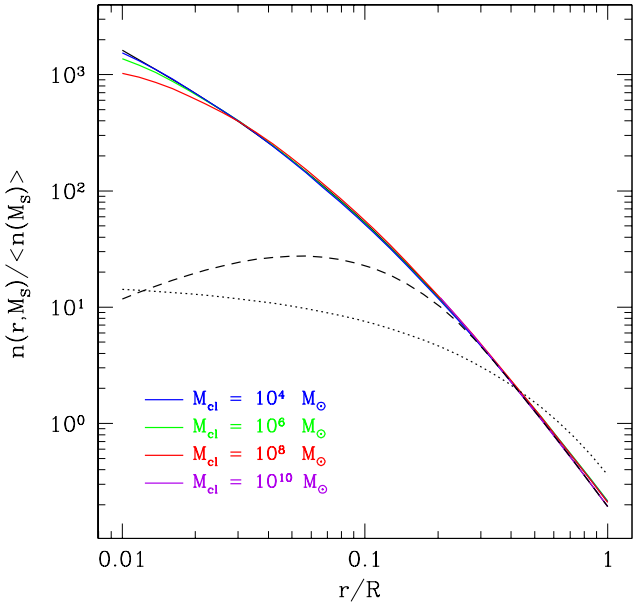}}
\caption{Theoretical non-truncated subhalo number density profiles per
  infinitesimal mass for subhaloes with $M\cl$, scaled to the total
  number, for subhaloes with different masses (solid coloured lines),
  compared to the halo mass density profile of the NFW profile (solid
  black line). The effects of including a primordial diffuse particle
  component that is progressively accreted by haloes is also shown
  (dashed coloured lines). For comparison, we also plot the Einasto
  law fitting the {\it truncated} subhalo number density profiles in
  simulated haloes (dotted black line).}
\label{f3}
\end{figure}

In Figure \ref{f3}, we show the theoretical number density profile per
infinitesimal mass for subhaloes with $M\cl$, $n(r,M\cl)\equiv
(4\pi r^2)^{-1} N(r,M\cl)$, scaled to the total mean number density of
such subhaloes, $\langle n(M\cl)\rangle$, so obtained. This scaled
number density profile shows a cutoff, preceded by a short bending, at
small enough radii that depends on the subhalo mass (see e.g. the
curves for subhaloes with $10^{8}$ \modot and $10^{6}$
\modotc)\footnote{The cutoff for subhaloes with $10^{10}$ \modot is
  located at $\log(r/R)\sim 0.1$, but it is not preceded by any short
  bending likely due to the effects mentioned above concerning the
  approximate conditional number density of nested peaks at large
  $\R$.}. The reason for this behaviour, also found in numerical
simulations (see \citealt{ALBF09}), is well-understood: there can be
no clump with mass $M\cl$ inside the radius $r$ encompassing that mass
because the accretion of such a subhalo at the time where the halo had
the mass $M(r)\sim M\cl$ would automatically cause a major merger and
the consequent destruction of the merging objects. Apart from that
short bending and cutoff, all the scaled number density profiles for
subhaloes with different masses overlap with the mass density profile
of the halo. The ratio between the subhalo number density and the
total halo mass density predicted by the model (see
eq.~[\ref{densclump}]),
\beq
\frac{n(r,M\cl)}{\rho(r)}\!=\!\!\left(\!\!\left\{\!\!\left[\frac{\der N_{\delta\pk}(M\cl)}
{\der \delta\pk}\!\right]_{\!\delta\pk(\R)}\!\!\!\frac{\der\delta\pk}{\der \R}\!\right\}_{\!\!\R(M)}\!\!
\frac{\der \R}{\der M}\!\right)_{\!\!M(r)}\!\!\!,
\label{ratio} 
\eeq
is flat, indeed. This result is at odds with that found in
simulations, where the scaled number density profiles for subhaloes of
different masses also overlap with each other, but show a much less
steep profile than the halo mass density profile (see the dotted curve in
Fig.~\ref{f3}).

The situation in simulations indicates that there is a diffuse dark
matter component outside subhaloes that becomes dominant at small
radii (SWV). But is this reasonable? In idealised hierarchical
cosmologies, all the dark matter is expected to be locked into
virialised haloes of different masses that develop through minor and
major mergers. And the same is true for the matter inside haloes: all
of it is expected to be locked into subhaloes of different
masses. Even if subhaloes are tidally truncated by the host potential
well (see Sec.~\ref{truncation}), the liberated matter will be in the
form of subhaloes previously seen as subsubhaloes. It is true that
when dark matter begins to cluster, after decoupling (or after
equality if decoupling took place earlier), it is in the form of a
diffuse component which is accreted by the first haloes formed by
monolithic collapse. But accretion of diffuse matter proceeds in a
very short time compared to the cosmic times we are interested in, so
we can neglect such a transient phase\footnote{This may not be the
  case for warm dark matter, whose decoupling marking the beginning of
  the clustering takes place much later.}. In simulations, it is
instead normal to find some amount of diffuse dark matter in current
haloes for two reasons: i) simulations start with unclustered dark
matter at much smaller redshifts (of about $100$) and ii) haloes
(subhaloes) below the resolution mass contribute to a melt diffuse
component until it is fully accreted by more massive haloes
(subhaloes). As a consequence, only about $60$ \% of the total mass in
current haloes is aggregated through minor and major mergers (about
$40$ and $20$ \%, respectively); all the remaining mass is accreted in
the form of diffuse dark matter \citep{Wanea11}. This modifies the
hierarchical way CDM haloes cluster at the small mass end and leads to
the presence of diffuse particles until quite a large $z$ \citep{AW10}. 
Therefore, if we are to compare the predictions of the
model with the results of \nbody simulations, we must account for this
effect.

In the presence of diffuse dark matter, the result above that the
scaled subhalo number density profile overlaps with the halo mass
density makes no sense. It would imply that the fraction of mass in
the form of the diffuse component has the same density profile or,
equivalently, that the mass fraction accreted by haloes in the form of
diffuse matter is constant, whereas the diffuse dark matter should be
progressively accreted, so its fraction falling into haloes should
diminish with increasing time and, given the inside-out growth of
accreting haloes, with increasing radius in any individual halo, 
causing a downward bending of the subhalo number density towards
the halo centre as observed in simulated haloes.

To calculate the expected bending of the subhalo number density at
small radii {\it in a given simulation}, we need to know the
time-evolving mass fraction in the diffuse component outside haloes,
$f\dc(t)$. This mass fraction satisfies the differential equation
\beqa
\frac{\der f\dc}{\der t}=\frac{f\dc(t)}{1-f\dc(t)}\,\frac{1}{\bar\rho(t)}\int_{M\res}^\infty
\der \widetilde M r\acc(\widetilde M,t)N(\widetilde M,t) \nonumber\\
\equiv \frac{f\dc(t)}{1-f\dc(t)}\,r\dc(t) \,,~~~~~~~~~~~~~~~~~~~~~~~~~~~~~~~~~~~~~
\label{feq}
\eeqa
where $r\acc(M,t)$ is the mass accretion rate of haloes with mass $M$
at $t$, given by equation (\ref{dmd}), and $M\res$ is the mass
resolution of the simulation. The solution of equation (\ref{feq}) for
the initial condition $f\dc(t\res)=1$, with $t\res$ the starting time
of the simulation, is given by the implicit equation
\beq
f\dc(t)-\ln[f\dc(t)]=1-\int_{t\res}^{t} \der \tilde t\,r\dc(\tilde t)\,.
\label{fdc}
\eeq
Thus, the mass accreted by the halo at $t$ in the form of subhaloes is
diminished by a factor $1-f\dc(t)$ compared to that in the case of no
primordial diffuse component. Given the halo inside-out growth, this
implies that the contribution from subhaloes to the halo mass density
profile changes from $\rho(r)$ to $\rho(r)\{1-f\dc[t(r)]\}$, with
$t(r)$ the time where the accreting halo reaches the radius $r$. The
effect of such a time varying mass fraction in the diffuse component
for the initial cosmic time $t\res$ corresponding to $z=127$ and the
resolution mass $M\res$ equal to $10^4$ \modot as in SWV simulations
is shown in Figure \ref{f3}. The curve so obtained is much like the
one found by SWV, although not identical. As we will see in Section
\ref{truncation}, the difference is likely due to the effects of
subhalo truncation not considered yet.

The presence of a diffuse dark matter component should have very
little effect, however, on the cumulative subhalo abundance, $N( >
M\cl)$, shown in Figures \ref{f1} and \ref{f2}. The reason is that,
despite the outward-decreasing subhalo number density profiles, the
number of subhaloes increases with radius, meaning that they are
mostly aggregated by the halo at late times when essentially all the
diffuse dark matter component has already disappeared (even in
simulations). Only if we were analysing the subhalo abundance at very
high redshifts (or very small radii) should the effect of the diffuse
dark matter component also be taken into account when dealing with the
subhalo abundance. Note that the same is true for the halo mass
function: at very high redshifts it will be affected by the diffuse
dark matter component, which should not be present in the real CDM
universe. This is not taken into account in studies of that quantity
from \nbody simulations.

To sum up, in the case of (essentially) no primordial diffuse dark
matter component, as in the real CDM universe, the scaled number
density profile for (non-truncated) subhaloes with any given mass
should coincide with the scaled halo mass density profile. On the
contrary, in the case of a substantial amount of diffuse dark matter,
as in numerical simulations or in the real universe soon after
decoupling (or after the time of equality), the scaled number
density profiles for subhaloes of different masses should also overlap
with each other, but not with the halo mass density profile. They should
be equal to this latter profile {\it times the factor $1-f\dc(r)$
  giving the mass fraction clustered in haloes by the time $t(r)$ when
  the halo reached the radius $r$}.

An important consequence of the previous result is that the spatial
distribution of (non-truncated) subhaloes is the same in haloes grown
by PA as in haloes having suffered major mergers. If it were different
in both kinds of haloes, then the typical mass density profile would
also be different, which would be in contradiction with the results of
\nbody simulations (see SVMS). Strictly, the possibility remains that
the deviation in the typical mass density profile for subhaloes of
some mass is exactly balanced by that for subhaloes of the remaining
masses, but such an arrangement is very unnatural. We therefore
conclude that the spatial distribution of subhaloes must be
independent of the host aggregation history. As discussed in SVMS,
this conclusion, far from being unexpected, reflects the fact that
virialisation is a real relaxation process. As such, it must cause the
memory loss of the initial conditions, not only regarding the smooth
halo structure and kinematics, but also regarding halo substructure
(but see Sec.~\ref{friction}).

\section{THE EFFECTS OF TRUNCATION}\label{truncation}

When subhaloes are aggregated by a halo, they are tidally truncated by
its potential well. Consequently, to compare the predictions of the
model with the results of numerical simulations we must account for
the effects of truncation. In fact, tidal truncation alters not only
the mass of subhaloes but also the number of subhaloes with a given
original non-truncated mass, $M\cl$, due to the appearance of new
first-level subhaloes of that mass, previously seen as subsubhaloes,
that are liberated from their host subhaloes.

Let us first concentrate in the change produced in {\it the number of
  subhaloes with a given non-truncated mass.} The number density per
infinitesimal mass of subhaloes with non-truncated mass $M\cl$
corrected for the number density effect owing to truncation is
\beqa 
N\tr(r,M\cl)=N(r,M\cl)+\int_{M\cl\mini}^M \der \widetilde M\cl\, N(r,\widetilde M\cl)\nonumber\\
\times\int_{R\tr(r,\widetilde M\cl)}^{R(\widetilde M\cl)} \der \tilde r\,N\tr_{\widetilde M\cl,M(r)}(\tilde r,M\cl)\,,~~
\label{corr} 
\eeqa
where $N(r,M\cl)$ is the number density of non-truncated subhaloes
with mass $M\cl$, calculated in Section \ref{numdens},
$R\tr(r,\widetilde M\cl)$ is the truncation radius of the original
subhaloes with mass $\widetilde M\cl$ located at $r$ and $M\cl\mini$
is the minimum subhalo mass that gives rise by truncation to new
first-level subhaloes with $M\cl$. The subindexes in the second-level
(differential) truncated subhalo abundance, $N\tr_{\widetilde
M\cl,M(r)}(\tilde r,M\cl)$, indicate that this subhalo number density
profile corrected for truncation refers to a host, in this case a
subhalo, with mass $\widetilde M\cl$ at the time when it was
aggregated by the halo, with a mass at that moment equal to $M(r)$,
and hence, different from the mass $M_0$ of the halo at $t_0$.

We will consider two extreme cases. In case (a), all CDM particles are
in subhaloes of a certain mass, as theoretically expected in the real
CDM universe at late times, so the truncation of first-level subhaloes yields
only new subhaloes previously seen as subsubhaloes; we thus have
$M\cl\mini=M\cl$. In case (b), subhaloes are instead essentially made
of diffuse particles, so the truncation of first-level subhaloes
does not modify the number of these subhaloes (just their mass as well
as the total mass of diffuse particles in the intrahalo medium); we
thus have $M\cl\mini=M$. Clearly, in case (b), equation (\ref{corr})
has the trivial solution $N\tr(r,M\cl)=N(r,M\cl)$, while in case (a)
equation (\ref{corr}) is an integral equation neither of
Fredholm nor of Volterra type, but can still be solved in the way
explained in Appendix \ref{A}.

\begin{figure}
\vskip -15pt
\centerline{\includegraphics[scale=0.46]{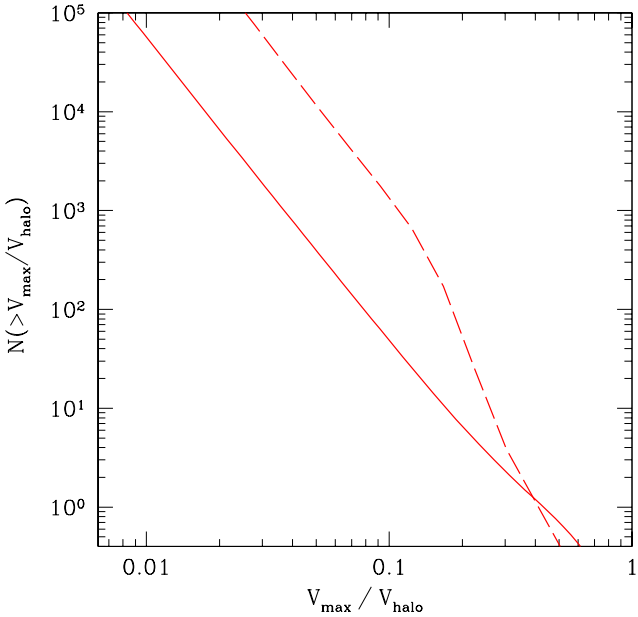}}
\caption{Same as Figure \ref{f1} but for truncated subhaloes in cases
(a) (dashed red line) and (b) (solid red line) corresponding to
subhaloes made of subsubhaloes (and so on) and of a diffuse particle
component, respectively. The cumulative abundance of truncated
subhaloes in case (a) coincides with the cumulative abundance of
non-truncated subhaloes shown in Figure \ref{f1}.}
\label{f4}
\end{figure}

The effect of truncation in the cumulative subhalo abundance as a
function of $V\maxi$ is shown in Figure \ref{f4}. Note that the
quantity $V\maxi$ is insensitive to the strength of tidal truncation
because the maximum circular velocity in a subhalo is reached at a
radius smaller than the truncation radius. This is the reason why
numerical studies usually plot the subhalo abundance as a function of
$V\maxi$ instead of as a function of the truncated mass much harder to
estimate (see the discussion below). In case (b), truncation does not
produce any apparent change in the cumulative subhalo abundance
relative to that plotted in Figure \ref{f1}, recovering that found in
simulations.  The reason is that subhaloes harbour only diffuse
particles, so the subhalo number does not change where subhaloes are
truncated. In contrast, we do expect an important change in case (a),
better suited for the real CDM universe, owing to the appearance of
new subhaloes previously seen as subsubhaloes. As shown in Figure
\ref{f4}, the abundance of first-level subhaloes then increases
dramatically (about two orders of magnitude) in comparison with the
abundance shown in Figure \ref{f1} and found in numerical simulations.

\begin{figure}
\vskip -15pt
\centerline{\includegraphics[scale=0.46]{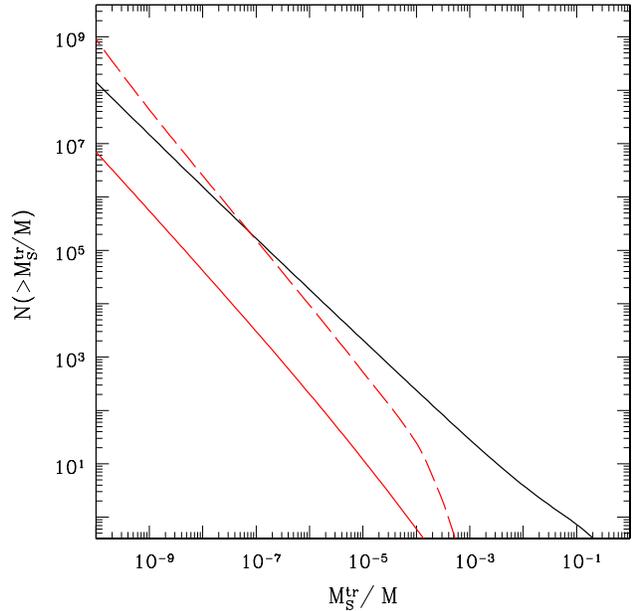}}
\caption{Same as Figure \ref{f4} but as a function of truncated mass
$M\cl\tr$. For comparison, cumulative abundance of non-truncated subhaloes 
(solid black line) as in Figure \ref{f1} but as a function of
non-truncated mass $M\cl$ instead of maximum circular velocity
$V\maxi$.}
\label{f5}
\end{figure}

All the previous results favouring case (b) indicate not only that, in
simulations, a large fraction of the mass in simulated haloes is in
the form of a diffuse component (see e.g. SWV), but also that such a
diffuse component must be widely dominant {\it in subhaloes} so that
very few new subhaloes emerge by tidal truncation of other
subhaloes. This does not necessarily mean that there is no subhalo at
any level higher than one. It just indicates that subsubhaloes must be
rare enough for not having significant effects on the general
properties of substructure as drawn from current high-resolution
numerical simulations. SWV report the detection of subhaloes up to
third level. However, according to the present results, these
third-level subhaloes should be seen only within the most massive
subhaloes and their most massive subsubhaloes, so that the total
number of subhaloes liberated by truncation would be insignificant
compared to the number of them directly aggregated. There are several
reasons for such an important lack of subsubhaloes in
simulations. Subsubhaloes are previously truncated by the
subhalo potential well and this is also true for third-level subhaloes
within subsubhaloes themselves and so on. The higher the level of
subhaloes, the earlier they typically form and the less massive they
typically are. The earlier subhaloes form, the larger is their mass
fraction below the mass resolution. And, the less massive the
subhaloes, the more centrally concentrated they are, so the more
severe the tidal disruption they yield in their own
subhaloes. Therefore, we expect the mass in simulated subhaloes to be,
indeed, mostly in the form of diffuse particles.

But this is not what we expect to find in the real CDM universe with
(essentially) no primordial diffuse component. Neglecting the minimum
halo mass, all haloes have grown through mergers between less massive
progenitors previously formed, so substructure should essentially obey
case (a). The difference between the first-level subhalo abundance
predicted by the model in cases (a) and (b) implies that there should
be, in the real CDM universe, {\it two orders of magnitude more
  first-level subhaloes than usually considered} from the results of
numerical simulations. This, together with the fact that in the real
CDM universe the subhalo number density profile is steeper than found
in simulations (see below), might have important implications on the
detectability of CDM from the enhanced flux of cosmic rays produced in
its annihilation in nearby subhaloes (e.g. \citealt{Spea08b,EWT09}). 
However, the abundance of {\it dwarf galaxies} in,
say, a Milky Way mass halo {\it is not affected} because the tidal
truncation of subhaloes does not liberate new galaxies that were
previously hidden. Moreover, luminous (and cold baryonic) matter
usually lies {\it at the centre} of subhaloes, so the subhalo
abundance relevant for the expected number of satellite galaxies
rather corresponds to case (b).

We can now turn to the second effect: the change in {\it the mass of
  the truncated subhaloes.} To express the preceding subhalo abundance
and number density profiles as a function of the subhalo truncated
mass $M\cl\tr$,\footnote{In simulations, the subhalo mass is usually
  taken equal to the truncated mass $M\tr\cl$ plus the unbound mass
  belonging to the halo in the volume occupied by the subhalo, denoted
  by $M\sub$. We have checked that the use of $M\sub$ instead of
  $M\tr\cl$ does not significantly alter the results presented here.}
we must take into account the relationship between that mass and the
original non-truncated mass $M\cl$,
\beq 
M\cl\tr(r,M\cl)= 4\pi\int_0^{R\tr(r,M\cl)} \der\tilde r\, \tilde r^2\,\rho_{M\cl, M(r)}(\tilde r)\,,
\label{mtr} 
\eeq
where $\rho_{M\cl,M(r)}(r)$ is the typical subhalo density profile
equal to that for haloes with mass $M\cl$ at the time of the subhalo
aggregation when the halo had a mass equal to $M(r)$. 

The cumulative abundances of truncated subhaloes for Milky Way mass
haloes as a function of $M\cl\tr$ are plotted in Figure \ref{f5}. The
log-log slopes we find are equal to $-1.12$ and $-1.05$ for cases (a)
and (b), respectively (or $-2.12$ and $-2.05$ for the {\it
  differential} subhalo abundance). The slope found by SWV in their
numerical simulations was $-0.90\pm 0.03$ ($-1.90 \pm 0.03$), hence
once again closer to the value predicted by the model in case
(b). Note that, although the difference between the theoretical and
empirical slopes in case (b) is small, it may be essential for having
a convergent or divergent number of subhaloes for masses approaching
to zero. It is true that this limit is actually not reached due to the
cutoff in the power-spectrum (and the non-negligible velocity
dispersion) of dark matter particles, but those slopes still tell at
which extent the mass fraction in low-mass subhaloes is dominant or
not. The possible origin of the slight departure in the slope
between the predictions of the model and the results of numerical
simulations is discussed below. In any case, even if the total number
of subhaloes in case (b) converged, that in case (a) should diverge as
found here, so our results point, in the case of CDM cosmologies, to a
halo mass fraction in the form of low-mass subhaloes (below the resolution
of current simulations) much larger than usually
thought. 

The subhalo number density per infinitesimal mass corrected for
truncation as a function of the truncated subhalo mass, $M\cl\tr$, or
simply the real truncated subhalo number density per infinitesimal
mass, takes the form
\beqa 
N\tr(r,M\cl\tr)=N[r,M\cl(r,M\cl\tr)]+\int_{M\cl\mini}^M \der \widetilde M\cl\, N(r,\widetilde M\cl)\nonumber\\
\times\int_{R\tr(r,\widetilde M\cl)}^{R(\widetilde M\cl)} \der \tilde r\,N\tr_{\widetilde M\cl,M(r)}(\tilde r,M\cl\tr)\,,~~~~~~~~~~~
\label{corr3} 
\eeqa
with the function $M\cl(r,M\cl\tr)$ implicitly defined by equation
(\ref{mtr}). The integral equation (\ref{corr3}) can be solved for
$N\tr(r,M\cl\tr)$ in the two extreme cases (a) and (b) above in the
same way as equation (\ref{corr}) for $N\tr(r,M\cl)$. Note that the
truncation radius $R\tr$ in equations (\ref{corr}) and (\ref{corr3})
depends, for a given halo mass, not only on the radius of the subhalo
at the aggregation time and its non-truncated mass, but also on its
orbit, which in turn depends on the host kinematics, modelled in SSMG.

\begin{figure*}
\vskip -125pt
\centerline{\includegraphics[scale=0.75]{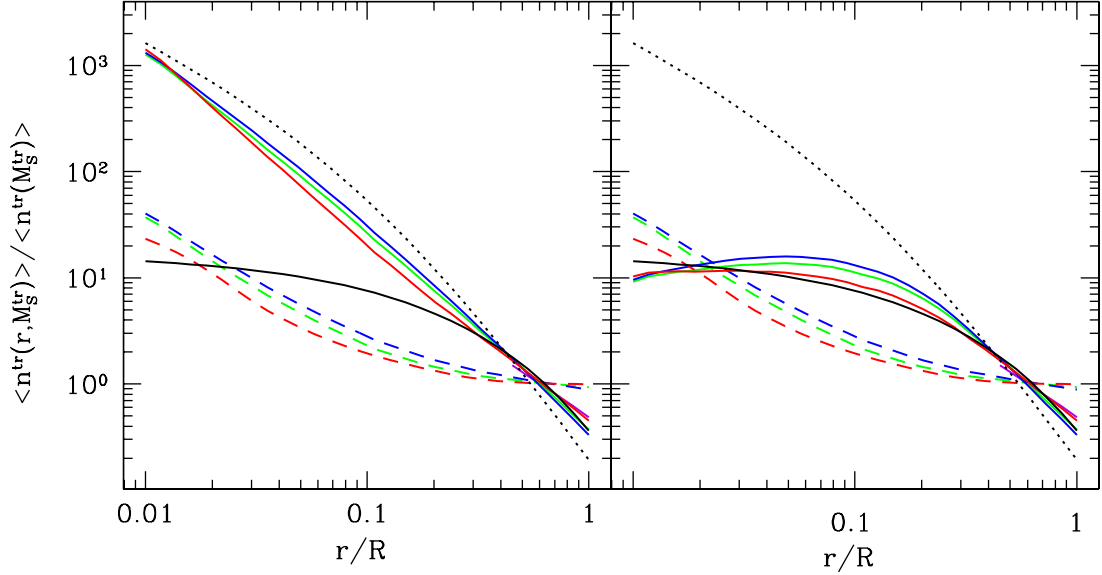}}
\vskip -25pt
\caption{Same as Figure \ref{f3} but for {\it truncated} subhalo
  masses $M\cl\tr$, in the same cases (a) (dashed lines) and (b)
  (solid lines) as in previous figures, compared now to the Einasto
  law (solid black line) that fits the subhalo number density profiles
  drawn from simulations (SWV). For comparison with Figure
  \ref{f3}, we also plot the NFW law fitting the predicted number
  density of non-truncated subhaloes (dotted black line). {\it Left
    panel:} predictions for the case of no primordial diffuse particle
  component.  {\it Right panel:} case (b) predictions for the case of
  a primordial diffuse particle component that is progressively
  accreted by haloes. }
\vskip -35pt
\label{f6}
\end{figure*}

\begin{figure*}
\vskip -125pt
\centerline{\includegraphics[scale=0.75]{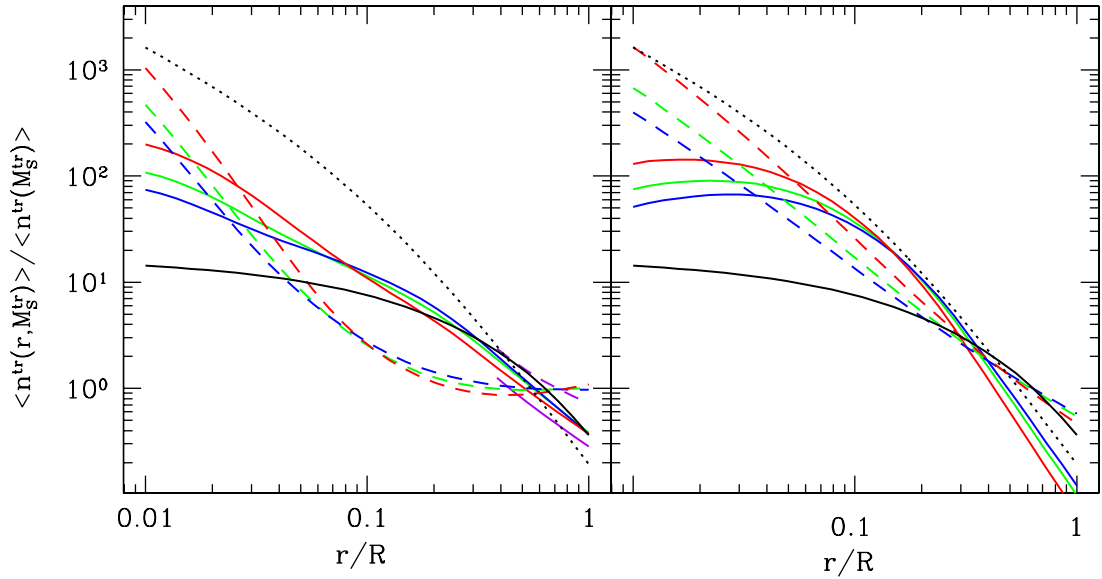}}
\vskip -25pt
\caption{Same as right panel of Figure \ref{f6} but using the \citet{ZJMB09} 
(left panel) and \citet{Kea11} (right panel) $M$-$c$
toy models instead of the SMGH physical model to calculate the effects
of subhalo truncation by the host potential well.}
\label{f7}
\end{figure*}

The truncation radius is very hard to determine in numerical
simulations. This is why different authors adopt different procedures
leading to somewhat different results. For instance, \citet{Dea07} 
took the radius at which the density of the subhalo
(corrected for the local background contribution) is equal to the
local background density, while SWV adopted the radius at which the
clump mean inner density (also corrected for the local background
contribution) is 0.02 times the mean inner background density, shown
to lead to a truncation radius in overall agreement with the
theoretical tidal radius defined by \citet{bt87}. In the
present model, we adopt the better motivated truncation radius given
in \citet{GMS94}. These authors showed that,
regardless of the shape of the orbit, subhaloes are truncated by the
host potential well essentially at the radius encompassing an inner
mean density equal to that of the host halo at the clump
pericentre. Note that such a truncation radius would roughly coincide
with Diemand et al.'s provided clumps described circular orbits;
unfortunately, this is not the case in general. On the other hand, it
would coincide with SWV. truncation radius provided the halo mean
inner density at the subhalo pericentre were 0.02 times the local halo
density, which is, in general, not the case either. Assuming subhaloes
with the median velocity for a normal distribution with radial and
tangential velocity dispersions given by the model in SSMG, we
determined the typical pericentre reached by subhaloes located at any
given radius. Then, assuming the non-truncated subhaloes at their
aggregation time with the typical halo density profile with the
mass-concentration ($M$--$c$) relation given by \citeauthor{smgh07} 
(\citeyear{smgh07}, hereafter SMGH\footnote{The $M$--$c$ relation provided by
  SMGH is consistent with the SVMS model for CDM haloes (see SVMS).}),
we calculated, from the halo mean inner density at the resulting
subhalo pericentre, the wanted subhalo truncation radius.

Figure \ref{f6} shows the theoretical number density profile per
infinitesimal mass for truncated subhaloes with $M\tr\cl$, scaled to
the total number of such subhaloes for Milky Way mass haloes in cases
(a) and (b) compared to those found in numerical simulations or, more
exactly, to the shallow profile of the Einasto form fitting them. In
numerical simulations, these scaled number density profiles are indeed
found to be much shallower than the halo density profile 
(\citealt{DMS04b,Gae04,NK05,Dea07}; SWV) and independent of subhalo mass 
(\citealt{DMS04a}; SWV; \citealt{Lud09}). In the left panel of Figure \ref{f6}, we see
that the truncated subhalo number density profiles predicted by the
model in case (a), with no primordial diffuse component, are on the
contrary similarly steep as the NFW profile fitting the profiles found
for non-truncated subhaloes and show now a slight dependence on
subhalo mass. The situation does not improve in case (b): the number
density profiles show a more marked dependence on subhalo mass and are
much shallower than found in simulations at large radii, while they
become steeper at small radii. And an intermediate case between (a)
and (b) would not improve the results: the theoretical number density
profiles would always show a clear dependence on $M\tr\cl$ and be
convex instead of concave.

As mentioned, the fact that the empirical subhalo number density
profiles are independent of subhalo mass and shallower than the halo
mass density profile implies that, in numerical simulations, a
substantial fraction of dark matter is in the form of a diffuse
component that increases inwards. According to the discussion in
Section \ref{numdens}, a bending of the theoretical number density
profiles towards those found in simulations is expected, indeed, in
the case that there is a primordial diffuse component that is
progressively accreted by haloes. Such an effect was calculated in
Section \ref{numdens} for non-truncated haloes. Thus, we have repeated
the same calculations for truncated haloes in case (b) (in case (a)
there is no diffuse component). That is, we have considered that the
contribution of subhaloes to the halo mass density is given by
$\rho(r)[1-f\dc(r)]$ instead of $\rho(r)$. Then, the log-log slope of
the truncated subhalo abundance does not essentially change, but the
truncated subhalo number density profiles do markedly. As shown in the
right panel of Figure \ref{f6}, they then become essentially in
agreement with the results of numerical simulations.

The small deviations that still remain (in the curvature of the
profiles and in the dependence on subhalo mass) are likely due to the
SMGH $M$--$c$ relation used in the modelling of truncation for very
large $z$, which apparently is not accurate enough. To see the kind of
effect the adoption of one particular $M$--$c$ relation has on these
results, we have repeated the same calculation with two different
$M$--$c$ relations: those provided by \citet{Kea11} and \citet{ZJMB09}. 
The predicted log-log slopes of the differential truncated
subhalo abundance are then equal to $-2.07$ and $-2.01$ using Klypin
et al. $M$--$c$ relation and $-2.04$ and $-2.01$ using Zhao et
al. $M$--$c$ relation for cases (a) and (b), respectively (to be
compared with the slopes of $-2.12$ and $-2.05$ found using the SMGH
$M$--$c$ relation and the slope of $-1.9$ found in numerical
simulations by SWV). The truncated subhalo number density profiles in
case (b) predicted using \citet{Kea11} and \citet{ZJMB09} 
$M$--$c$ relations are shown in Figure \ref{f7}. As can be seen, the
theoretical subhalo number density profiles so obtained deviate more
markedly from the profiles drawn from simulations and are more
mass-dependent than those shown in the right panel of Figure
\ref{f6}. Therefore, the predictions drawn from the SMGH $M$--$c$
relation are neatly preferable. It may seem strange that the $M$--$c$
relations drawn from numerical simulations give poorer results than
those derived from the SMGH model. We note, however, that 
\citet{Kea11} and \citet{ZJMB09} $M$--$c$ relations do {\it not}
actually fit the results of simulations; they are the extrapolations
of the real fitting expressions, through some guessed toy models, to
the much wider mass and $z$ domains involved in the present
calculations. In any event, these results clearly show that any slight
deviation of the $M$--$c$ from the true relation has notable effects,
indeed, in the theoretical truncated subhalo number density profile.

Before concluding this Section, it is important to remark that, as the
non-truncated subhalo number density profile, $N(r,M\cl\tr)$, is
independent of the halo aggregation history (see Sec.~\ref{numdens})
and so are also both the anisotropy and velocity dispersion profiles
(SSMG) and the mass density profiles (SVMS) setting the truncation
radii, the truncated subhalo number density profiles,
$N\tr(r,M\cl\tr)$, must be also independent of the halo aggregation
history. In other words, all the properties derived so far should not
depend on the halo aggregation history. This would explain why all
substructure properties derived from numerical simulations reporting
to haloes with very different aggregation histories, show such small
scatters.

\section{THE EFFECTS OF DYNAMICAL FRICTION}\label{friction}

But things are not that simple. The spatial distribution of subhaloes
is also affected by dynamical friction resulting from gravitational
two-body interactions between subhaloes themselves and between subhaloes and
diffuse dark matter particles. As in a major merger, the radial
location of subhaloes suffers an important scrambling, the effects of
dynamical friction that have previously taken place are essentially
erased. Therefore, the effects of dynamical friction depend on the
time elapsed since the last major merger. This means that, contrarily
to all the processes previously mentioned\footnote{The radial mapping
  of haloes is only preserved during accretion periods. However, the
  structural and kinematic properties of haloes resulting from a major
  merger are indistinguishable from those shown by haloes grown by PA
  (see SVMS and SSMG).}, dynamical friction may lead to significant
differences between haloes according to their aggregation history
(through the time elapsed since the last major merger).

The large number of tiny subhaloes and of diffuse dark matter (in
numerical simulations) suggests that dynamical friction should be very
effective, at least in the case of the most massive subhaloes more
prone to suffer it. But, this should have repercussions on the smooth
structure and kinematics of haloes, which should then depend on the
halo aggregation history, while there is no clear sign of such a
dependence in simulations (see SVMS and references
therein). Furthermore, as a consequence of dynamical friction, the
most massive subhaloes should lie closer to the halo centre than less
massive ones, whereas subhaloes of different masses show identical
scaled number density profiles. And the only minor difference is
rather of the opposite sign: the more massive the subhalo, the larger
the minimum radius reached by its scaled number density profile
\citep{ALBF09}. In particular, there is no sign of very massive
haloes being accumulated at the halo centre. Therefore, simulated
haloes show no apparent effect of dynamical friction.

The only way to escape this paradox seems to be that the effects of
dynamical friction may be present but go unnoticed. Does this make
sense? If the only subhaloes having had time to suffer significant
dynamical friction since the last major merger were the most massive
ones, it could be very difficult to detect it because the number of
those subhaloes is so small that their number density profile is
quite uncertain (due to large Poisson errors). Certainly, they could
be quite numerous at the halo centre where they should tend to
accumulate, but, when a very massive subhalo falls to the halo centre
and merges with the massive subhalo already lying there or simply
settles down well-centred, it becomes invisible because it mimics the 
central part of the host halo. This effect
could still manifest itself through an increased amount of small
subhaloes near the halo centre, corresponding to old second-level
subhaloes (in the massive subhaloes having fallen to the halo
centre) converted to first-level ones. But such an effect should only
be observable in case (a), because, in case (b), subhaloes are mainly
made of diffuse dark matter. In this sense, the lack of such an
indirect proof of dynamical friction would be an additional argument
in favour of case (b) when trying to model the results of \nbody
simulations.

\section{SUMMARY AND CONCLUSIONS}\label{dis}

The evolution of nested peaks in the filtering of the primordial
density field truthfully traces the evolution of halo substructure
developing at all levels as a consequence of halo accretion and major
mergers. Thus, the peak formalism can be used to describe subhalo
abundance in typical haloes. Moreover, taking into account that haloes
growing by PA develop from the inside out, it can be used to derive
the non-truncated subhalo number density profiles per infinitesimal
mass for subhaloes of different masses and, making use of the
halo structural and kinematic profiles modelled in SVMS and SSMG, one
can correct those quantities from tidal truncation as well.

The subhalo properties predicted in the $\Lambda$CDM cosmology for
Milky Way mass haloes are in very good agreement with those found in
numerical simulations provided dark matter within subhaloes is
essentially in the form of diffuse particles. The only slight
deviations found, in this case, between the theoretical predictions
and the results of numerical simulations seem to be due to the
non-fully accurate (sub)halo $M$--$c$ relation used. More accurate
$M$--$c$ relations drawn e.g. from the SVMS model of halo structure
would be welcome.

But the true subhalo properties expected on pure theoretical grounds
in a real CDM universe rather correspond to those predicted under the
opposite assumption that all dark matter in subhaloes is locked in
higher-order level subhaloes. Accurate predictions are also given for
this more realistic case. The most striking result is that there
should be, in this case, two orders of magnitude more subhaloes than
usually thought on the basis of the results of \nbody
simulations. This might have important implications for the
detectability of CDM, but it does not affect the dwarf galaxy
abundance estimated from CDM simulations.

In any of these two scenarios, subhalo properties are expected to be
independent of the halo aggregation history. This means that, despite
having been derived under the PA condition, all the previous
quantities should hold for haloes having suffered major mergers. The
only effects that could depend on the halo aggregation history are
those due to dynamical friction. According to the present results and
those found in SVMS and SSMG, dynamical friction can be neglected as
long as we are interested in modelling haloes in current
simulations. However, it might have visible effects in the real CDM
universe as well as in future, higher resolution, \nbody simulations.
Of course, dynamical friction is also expected to have important
consequences in baryon physics ignored in the present study.

\vspace{0.75cm} \par\noindent
{\bf ACKNOWLEDGEMENTS} \par

\noindent This work was supported by the Spanish DGES,
AYA2006-15492-C03-03 and AYA2009-12792-C03-01, and the Catalan DIUE,
2009SGR00217. One of us, SS, was beneficiary of a grant from the
Institut d'Estudis Espacials de Catalunya.

\clearpage

\appendix

\onecolumn 

\section{TRUNCATED SUBHALO NUMBER DENSITY IN CASE (\lowercase{a})}\label{A}

In case (a), i.e. $M\cl\mini=M\cl$, the subhalo number density per
infinitesimal mass corrected for truncation (eq.~[\ref{corr}]) takes
the form
\beq
N\tr(r,M\cl)=N(r,M\cl)+\int_{M\cl}^M \der \widetilde M\cl\, N(r,\widetilde M\cl)
\int_{R\tr(r,\widetilde M\cl)}^{R(\widetilde M\cl)} \der \tilde r\,N\tr_{\widetilde M\cl,M(r)}(\tilde r,M\cl)\,.
\label{corr0} 
\eeq
By partial integration, equation (\ref{corr0}) leads to
\beq
N(r,M\cl)=N(r,M\cl)+\int_{M\cl}^M \der \widetilde M\cl\, N(r,>\widetilde
M\cl)\frac{\der}{\der \widetilde M\cl} \int_{R\tr(r,\widetilde M\cl)}^{R(\widetilde M\cl)} \der \tilde r\,N\tr_{\widetilde M\cl,M(r)}(\tilde r,M\cl)\,.
\label{corr00} 
\eeq
Taking into account that both expressions (\ref{corr0}) and
(\ref{corr00}) hold for any arbitrary value of $M\cl$, we are led 
to\footnote{This is a physical rather than mathematical
implication. The dependence on $M\cl$ in the integrands does not allow
one to strictly prove the equality. But only a very unlikely
conspiracy would make it possible to balance any arbitrary change in
the integration limits by that produced in the integrands if they were
not really equal.}
\beq 
\frac{\der \ln}{\der\widetilde M\cl} [N(r,<\widetilde
M\cl)]=\frac{\der \ln}{\der \widetilde M\cl}
\int_{R\tr(r,\widetilde M\cl)}^{R(\widetilde M\cl)} \der 
\tilde r\,N\tr_{\widetilde M\cl,M(r)}(\tilde r,M\cl)\,,
\label{neweq} 
\eeq
implying 
\beq 
\T(r,M\cl)N(r,<\widetilde M\cl)= \int_{R\tr(r,\widetilde M\cl)}^{R(\widetilde M\cl)}
\der \tilde r \,N\tr_{\widetilde M\cl,M(r)}(\tilde r, M\cl)\,,
\label{neweq2} 
\eeq
where the function $\T(r,M\cl)$ is the unknown integration
constant. Choosing $\widetilde M\cl$ equal to $M(r)$ and taking into
account that haloes grow inside-out, the double subindex ``$\widetilde M\cl,
M(r)$'' in the truncated subhalo number density in the
integrand on the right can be chosen equal to ``$M_0,M_0$'' without any
loss of generality (see the meaning of such a double subindex in eq.~[\ref{corr}]). That is, for that particular value of $\widetilde
M\cl$, the subhalo is a clone of the host halo, except for the fact
that it has not grown since it was aggregated by the host halo. In
particular, its (sub)subhalo number density corrected for truncation
is identical to that of the host halo itself (the subindex ``$M_0,M_0$'' can
be omitted) and the (sub)subhalo with mass $M\cl$ is found at the same
minimum radius $r(M\cl)$ as in the host halo. Consequently, equation
(\ref{neweq2}) takes the form
\beq 
\T(r,M\cl)N(r)= \int^{r}_{{\rm max}\{r(M\cl),R\tr[r,M(r)]\}}
\der \tilde r \,N\tr(\tilde r, M\cl)\,,
\label{neweq3} 
\eeq
where we have taken into account that the $N[r,<M(r)]$ is but the
total subhalo number density at $r$, denoted as $N(r)$. 

At small $r$, we have $r(M\cl)\ge R\tr[r,M(r)]$ and differentiation of
equation (\ref{neweq3}) leads to
\beq
N\tr(r,M\cl)=\frac{\der}{\der r}\left[\T(r,M\cl)N(r)\right]\,.
\label{sol0}
\eeq
Substituting $N\tr(r,M\cl)$ given by equation (\ref{sol0}) into
equation (\ref{corr0}), taking into account equation (\ref{neweq2})
and the partial integration of the $\widetilde M\cl$-integral in the
resulting expression, we arrive at
\beq
\frac{\der}{\der r}\left[\T(r,M\cl)N(r)\right]=N(r,M\cl)+\frac{1}{2}\T(r,M\cl)
N(r)\,N(r,>M\cl)\,\left[2-\frac{N(r,>M\cl)}{N(r)}\right]\,.
\label{eqdif}
\eeq
For any reasonable (large enough) value of $M\cl$, $N(r,>M\cl)/N(r)$
is negligible in front of unity, so equation (\ref{eqdif}) takes the
simple form
\beq
\frac{\der}{\der r}\left[\T(r,M\cl)N(r)\right]=N(r,M\cl)+\left[\T(r,M\cl)
N(r)\right]\,N(r,>M\cl)\,.
\label{eqdifbis}
\eeq
This is a differential equation for $\T(r,M\cl)$, which can be solved
for the initial condition $\T[r(M\cl),M\cl]=0$ implied by equation
(\ref{neweq3}). Then, replacing the solution $\T(r,M\cl)$ into
equation (\ref{sol0}), we obtain the wanted number density per
infinitesimal mass of truncated subhaloes, $N\tr(r,M\cl)$.

At a large enough $r$, hereafter denoted by $r\en$, the condition
$r(M\cl) < R\tr[r,M(r)]$ will be finally met and this solution will
no longer hold. In this new regime, differentiation of equation
(\ref{neweq3}) leads to
\beq 
N\tr(r,M\cl)=\frac{\der}{\der
r}\left[\T(r,M\cl)N(r)\right]-\frac{\der R\tr[r,M(r)]}{\der
r}\,N\tr\{R\tr[r,M(r)],M\cl\}\,.
\label{sol1}
\eeq
Substituting $N\tr(r,M\cl)$ given by equation (\ref{sol1}) into
equation (\ref{corr0}), taking into account equation (\ref{neweq2})
and integrating by parts the integral over $\widetilde M\cl$ in the
resulting expression, we obtain
\beqa
\frac{\der}{\der r}\left[\T(r,M\cl)N(r)\right]+\frac{\der R\tr[r,M(R)]}{\der r}\,N\tr\{R\tr[r,M(r)],M\cl\}~~~~~~~~~~~~~~~~~~~~~~~~~\nonumber\\
=N(r,M\cl)+\frac{1}{2}\,\T(r,M\cl)\,N(r)\,N(r,>M\cl)\,\left[2-\frac{N(r,>M\cl)}{N(r)}\right]\,
\label{eqdif2}
\eeqa
which, for any reasonable (large enough) value of $M\cl$, reduces to
\beqa
\frac{\der}{\der r}\left[\T(r,M\cl)N(r)\right]+\frac{\der
R\tr[r,M(R)]}{\der r}\,N\tr\{R\tr[r,M(r)],M\cl\}\nonumber\\
=N(r,M\cl)+\left[\T(r,M\cl) N(r)\right]\,N(r,>M\cl)\,.
\label{eqdif2bis}
\eeqa

As $R\tr[r,M(r)]$ is smaller than $r$, the function
$N\tr\{R\tr[r,M(r)],M\cl\}$ has been previously obtained in the range
of small $r$, so the differential equation (\ref{eqdif2bis}) can then
also be solved for the function $\T(r,M\cl)$ with the initial
condition given by the value of $\T(r,M\cl)$ at $r\en$. Once
$\T(r,M\cl)$ has been determined, we can replace it in equation
(\ref{sol1}) to obtain the wanted function $N\tr(r,M\cl)$ in the new
radial range.

In fact, given that $r(M\cl)$ is greater than
$R\tr\{R\tr(r,M\cl),M[R\tr(r,M\cl)]\}$ in the relevant subhalo mass
range (i.e. except for $M\cl \la 10^3$ \modotc), the differential
equation (\ref{eqdif2bis}) can be solved analytically. Indeed, equation
(\ref{neweq2}) for $r=R\tr(r,M\cl)$ then takes the form
\beq 
\T[R\tr(r,M\cl),M\cl)]N(r)= \int^{R\tr(r,M\cl)}_{r(M\cl)}
\der \tilde r \,N\tr(\tilde r, M\cl)\,.
\label{neweq3bis} 
\eeq
Thus, by differentiating it with respect to $R\tr(r,M\cl)$ and
replacing the resulting expression for $N\tr\{R\tr[r,M(r)],M\cl\}$
into equation (\ref{eqdif2bis}), we arrive at
\beqa
\frac{\der \ln}{\der \ln r}\left[\T(r,M\cl)N(r)\right]=r 
\left\{1-\frac{1-N(r,M\cl)/\left[\T(r,M\cl)N(r)\right]}{1+N(r,>M\cl)}\right\}\,.
\label{eqdif5}
\eeqa
$N(r,M\cl)$ is much smaller than $N(r)$ and $\R(r,M\cl)$ is much
greater than one in this radial range, except for $M\cl\ga 10^{10}$
\modotc, as is checked a posteriori from equation
(\ref{neweq3}). Consequently, for $M\cl < 10^{10}$ \modot we can neglect
the term $N(r,M\cl)/[\T(r,M\cl)N(r)]$ on the right of equation
(\ref{eqdif5}), which leads to the following quite accurate solution
\beqa
\T(r,M\cl)N(r)=A(r\en)\,\exp\left[\int_{r\en}^r 
\der \tilde r\,\,\frac{N(\tilde r,>M\cl)}{1+N(\tilde r,>M\cl)}\right]\,,
\label{eqdif6}
\eeqa
with $A(r\en)$ an integration constant whose value is obtained by
continuity of the solution at $r\en$. Finally, differentiating
equation (\ref{eqdif6}) and taking into account equation
(\ref{neweq2}) we are led to
\beq
N\tr(r,M\cl)=\frac{A(r\en)\,N(r,>M\cl)}{1+N(r,>M\cl)}\exp\left[\int_{r\en}^r 
\der \tilde r\,\,\frac{N(\tilde r,>M\cl)}{1+N(\tilde r,>M\cl)}\right]\,.
\eeq
%


\end{document}